\documentclass[a4paper,12pt, epsfig]{article}
\pdfoutput=1
\usepackage{epsfig}
\usepackage{epstopdf}
\usepackage{graphicx}

\usepackage{amsmath}
\usepackage[psamsfonts]{amssymb}
\usepackage{euscript}

\usepackage{latexsym}

\renewcommand{\(}{\begin{equation}}
\renewcommand{\)}{end{equation} \vspace{-.05in}\linebreak}

\newcounter{saveeqn}
\newcounter{savealpheqn}

\newcommand{\alpheqn}{\setcounter{saveeqn}{\value{equation}}%
  \stepcounter{saveeqn}\setcounter{equation}{0}%
  \renewcommand{\theequation}{\mbox{\arabic{section}.\arabic{saveeqn}
\alph{equation}}}
  \renewcommand{\)}{\end{equation}}}
\def\part#1{\frac{\partial}{\partial{#1}}}%
\def\group#1{\refstepcounter{equation}\setcounter{saveeqn}
 {\value{equation}}%
  \label{#1}\setcounter{equation}{0}%
\renewcommand{\theequation}{\mbox{\arabic{section}.\arabic{saveeqn}
\alph{equation}}}
  \renewcommand{\)}{\end{equation}}}
\newcommand{\reseteqn}{\setcounter{equation}{\value{saveeqn}}%
  \renewcommand{\theequation}{\arabic{section}.\arabic{equation}}%
  \renewcommand{\)}{\end{equation}}}

\newcommand{\aalpheqn}{\setcounter{saveeqn}{\value{equation}}%
  \stepcounter{saveeqn}\setcounter{equation}{0}%
  \renewcommand{\theequation}{\mbox{
        \Alph{subsection}.\arabic{saveeqn}\alph{equation}}}
   \renewcommand{\)}{\end{equation}}}
\newcommand{\areseteqn}{\setcounter{equation}{\value{saveeqn}}%
  \renewcommand{\theequation}{\Alph{subsection}.\arabic{equation}}%
  \renewcommand{\)}{\end{equation}}}

\renewcommand{\thefootnote}{\alph{footnote}}
\renewcommand{\(}{\begin{equation}}
\renewcommand{\)}{\end{equation}}
\newcommand{\ba}{\begin{eqnarray}}
\newcommand{\ea}{\end{eqnarray}}

\newcommand{\bp}{\mathop{\vtop{\ialign{##\crcr
   $\hfil\displaystyle{}\hfil$\crcr\noalign{\kern-13pt\nointerlineskip}
   \BIG{(}\hskip0pt\crcr\noalign{\kern3pt}}}}}
\newcommand{\cbp}{\mathop{\vtop{\ialign{##\crcr
   $\hfil\displaystyle{}\hfil$\crcr\noalign{\kern-13pt\nointerlineskip}
   \BIG{)}\hskip0pt\crcr\noalign{\kern3pt}}}}}
\newcommand{\pa}{\mathop{\vtop{\ialign{##\crcr
    
$\hfil\displaystyle{\oplus}\hfil$\crcr\noalign{\kern+1pt\nointerlineskip 
}
   \hspace{.08in}$^{\alpha=0}$\hskip6pt\crcr\noalign{\kern3pt}}}}}

\newcommand{\beq}{\begin{equation}}
\newcommand{\eeq}{\end{equation}}

\numberwithin{equation}{section}

\catcode`\@=11
\def\vereq#1#2{\lower3pt\vbox{\baselineskip1.5pt \lineskip1.5pt
\ialign{$\m@th#1\hfill##\hfil$\crcr#2\crcr\sim\crcr}}}
\catcode`\@=12

\makeatletter
\newcommand\figcaption{\def\@captype{figure}\caption}
\newcommand\tabcaption{\def\@captype{table}\caption}
\makeatother
\renewcommand{\(}{\begin{equation}}
\renewcommand{\)}{\end{equation}}

\renewcommand{\beq}{\begin{equation}}
\renewcommand{\eeq}{\end{equation}}
\newcommand{\bea}{\begin{eqnarray}}
\newcommand{\eea}{\end{eqnarray}}
\newcommand{\beas}{\begin{eqnarray*}}
\newcommand{\eeas}{\end{eqnarray*}}

\newcommand{\bquo}{\begin{quote}}
\newcommand{\enqu}{\end{quote}}
\def\wav{\ensuremath{w_{\rm av}}}
\def\rd{\ensuremath{\rho_{\rm DE}}}

\begin{document}
\def\thefootnote{\fnsymbol{footnote}}

\def\thefootnote{\fnsymbol{footnote}}

\begin{center}
{\large {\bf
Model-Independent Dark Energy Equation of State\\from Unanchored Baryon Acoustic Oscillations  } }

\bigskip

\bigskip

{\large \noindent   Jarah Evslin\footnote{\texttt{jarah@impcas.ac.cn}}}

\end{center}

\renewcommand{\thefootnote}{\arabic{footnote}}

\vskip.7cm

\begin{center}
\vspace{0em} {\em  
  Institute of Modern Physics, CAS, NanChangLu 509, Lanzhou 730000, China\\}
\end{center}



\begin{abstract}
\noindent

\noindent
Ratios of line of sight baryon acoustic oscillation (BAO) peaks at two redshifts only depend upon the average dark energy equation of states between those redshifts, as the dependence on anchors such as the BAO scale or the Hubble constant is canceled in a ratio.  As a result, BAO {\it ratios} provide a probe of dark energy which is independent of both the cosmic distance ladder and the early evolution of universe.   In this note, we use ratios to demonstrate that the known tension between the Lyman alpha forest BAO measurement and other probes arises entirely from recent ($0.57<z<2.34$) cosmological expansion.  Using ratios of the line of sight Lyman alpha forest and BOSS CMASS BAO scales, we show that there is already more than 3$\sigma$ tension with the standard $\Lambda$CDM cosmological model which implies that either (i) The BOSS Lyman alpha forest measurement of the Hubble parameter was too low as a result of a statistical fluctuation or systematic error or else (ii) the dark energy equation of state falls steeply at high redshift.


\end{abstract}

%
\setcounter{footnote}{0}
\renewcommand{\thefootnote}{\arabic{footnote}}



\section{Motivation}

The location of the baryon acoustic oscillation peak provides a standard ruler which can and has been used to measure the expansion of the Universe.  This ruler can be observed along both the angular and line of sight directions.  However limited statistics implied that, until recently, precise measurements of this peak were only available for a weighted average of the line of sight and angular directions.  Using such a weighted average, studies typically found good agreement with the standard $\Lambda$CDM cosmological model.

This situation has changed with the separate measurements of the line of sight and angular peaks at $z=2.3$ by the BOSS survey \cite{1404.1801}, which is in mild tension with $\Lambda$CDM predictions when combined with data from other probes.  There have been numerous investigations of this tension, and distinct proposals for its resolution, but it is unclear just which of these should be chosen \cite{1411.1074}.

The goal of the present paper is quite modest.  Very recently the BOSS collaboration has released precise measurements of the line of sight and angular baryon acoustic oscillation peaks at low redshifts \cite{1509.0637}.  We will use this new data to show that the tension, should it be confirmed by future observations, arises entirely from the acceleration of the Universe between $z=0.57$ and $z=2.34$, thus eliminating many of the possible sources of the discrepancy suggested in earlier papers.

Our demonstration will be very elementary but also very model independent, and in fact entirely independent of the history of the Universe before $z=2.34$.  It will be based on a ratio of Hubble parameters arising from a ratio of BAO scales.  Such ratios have been considered in the past in similar contexts, although in general with additional assumptions, for example Ref.~\cite{1105.3838} assumes that the Universe never accelerated.  We make no assumptions about either the expansion history nor about the functional form of the dark energy equation of state.  We stress that such a general analysis is only possible now as a result of the precise anisotropic baryon acoustic oscillation measurement in Ref.~\cite{1509.0637}, indeed the larger uncertainties in older data implied that similar analyses revealed no tension \cite{busca2012}.

\section{Baryon Acoustic Oscillations and Model Dependence}
The spatial two-point correlation function of the density of baryons has a peak, the baryonic acoustic oscillation (BAO) peak, at a comoving scale $r_s$ which is believed to be about 150 Mpc.  As baryons on these scales have been nonrelativistic since shortly after recombination, the location of the peak in comoving coordinates has not changed.  The location of the peak therefore provides a universal ruler, with a constant comoving length at distinct redshifts through nearly all of cosmic history~\cite{baoteor}.   

Correlations may be observed for objects separated along the line of sight, whose distances are determined by redshifts $z$, or by objects separated perpendicular to the line of sight, whose distances are determined using their angular separation.   In this note we will be interested in the first case.  The two-point function in redshifts has a peak at
\beq
\Delta z = \frac{r_s H(z)}{c}
\eeq
where $H(z)$ is the Hubble parameter at redshift $z$.  Therefore BAO surveys in principle can determine the combination $r_s H(z)$ for various redshifts $z$.  

In practice, surveys accumulate data over a range of redshifts and package their results in terms of just one redshift for each sample.  This packaging requires the assumption of a fiducial cosmological model, however the dependence on the choice of model is quite small.  Similarly the position of the peak is determined by comparing the matter correlation functions with simulations based on a fiducial cosmological model, but due to nuisance parameters included in this analysis, the result is again quite robust with respect to changes in the fiducial model.  With these caveats understood, the resulting determination of $r_s H(z)$ is independent of the assumed cosmological model.  

On the other hand, the value of $r_s$ does depend on the cosmological model.  For example, as has been stressed in Ref.~\cite{14111094}, while the 2013 Planck results \cite{planck2013}, combined with polarization from WMAP, report a measurement of $r_s$ with an uncertainty of only 0.4\% assuming a standard $\Lambda$CDM cosmology, this uncertainty increases to 2.3\% if one modifies $\Lambda$CDM only by letting $N_{\rm eff}$, the effective number of light degrees of freedom, float freely.  Furthermore, fixing $N_{\rm eff}$ to the $\Lambda$CDM value leads to a $2.7\%$ shift in the central value of $r_s$.  While this model dependence is not large, it is already larger than the uncertainty obtained by some BAO measurements.

More generally, there are two ways in which $r_s$ may be modified.  First, one may fix the sound horizon size at recombination, fixing the locations of the acoustic peaks in the cosmic microwave background (CMB) power spectrum, but use an unconventional evolution of the sound horizon during the drag epoch such as the analysis of the streaming of supersonic baryons in Ref.~\cite{151003554}.  Second, one may modify the sound horizon at recombination, compensating for the shift in the angular size of the CMB acoustic peaks by modifying, for example, the evolution of dark energy at recent times to yield an angular diameter distance to recombination which changes proportionally $r_s$.

A modification of the acoustic horizon size at recombination can be achieved in two distinct ways, one can either modify the pre-recombination expansion $a(t)$ or else one may modify the speed of sound in the primordial plasma.  An exotic cosmological model may do either of these.  As an example of the first, note that standard inflationary cosmology asserts that the energy density of the Universe was twice dominated by dark energy, with no explanation as to its nature.  A third epoch of dark energy, well before matter-radiation equality, with negative energy density could lead to a brief stall in the expansion and so yield an increase in $r_s$.   As an example of the second mechanism, one may add charged matter in equilibrium with the plasma and with a density which is comparable to or even exceeds that of baryonic matter.  As the speed of sound in the plasma is inversely proportional to $\sqrt{3+R}$ where $R$ is the ratio of the energy density of charged matter to photons, this would increase $R$ and so decrease the speed of sound and so the sound horizon size.  Ordinarily such matter could be excluded by comparing the heights of the even and odd acoustic peaks in the CMB power spectrum.  However such a contribution could be minimized if the additional matter component is unstable and decays sufficiently before matter-radiation equality.  In a yet more extreme model, one may not assert that it decays, but rather adjust the primordial fluctuation spectrum to compensate for this effect.  All of these modifications (except for the freely floating $N_{\rm eff}$) are rather unnatural, but they serve to highlight that the precision with which $r_s$ is thought to be known results not from a direct measurement, but rather from the combination of a measurement with a wide array of assumptions which are yet to be tested.  


\section{Ratios of BAO Measurements}

Fortunately it is possible to use the radial BAO peak without knowing $r_s$.  If one knows the location of the peak at two different redshifts $z_1$ and $z_2$, then one obtains $r_s H(z_1)/c$ and $r_s H(z_2)/c$.  While each individually depends on the cosmological model through $r_s$, the ratio only depends on the expansion history in the time since these measurements.   Ratios of the tangential BAO peak similarly yield robust determinations of ratios of angular distances, which depend on integrals of $1/H(z)$ and the spatial curvature, however in this note we will not use them.  Combining ratios of angular and line of sight BAO measurements is equivalent to using only ratios of line of sight measurements plus Alcock-Paczynski tests \cite{aptest} on the BAO scale at each redshift.

We will use the final results from the Baryon Oscillation Spectroscopic Survey (BOSS) \cite{1509.0637} which provide measurements of $H(z)$ for samples of galaxies in two redshift groups.  The closer galaxy sample, called LOWZ, has an effective redshift of $z=0.32$ while the farther CMASS sample has an effective redshift of $z=0.57$.  These results appear quite consistent with the standard $\Lambda$CDM paradigm.  However we will also use BOSS measurements of the BAO in the autocorrelation of masses traced by the Ly$\alpha$ forest absorption of light from quasars \cite{1404.1801}  and the cross-correlation of the mass densities traced by the Ly$\alpha$ forest and quasars \cite{fontribera}.   These determine $H(z)$ at an effective redshift of $z=2.34$.  The autocorrelation and cross-correlation results for $H(z=2.34)$ were already combined in Ref.~\cite{1404.1801}. These results are all summarized in Table~\ref{htab}.  A number of other BAO measurements are not included in our analysis either because they do not decompose the BAO size into a line of sight and tangential component and/or because their survey volume overlaps with that of BOSS.

\begin{table}
\centering
\begin{tabular}{c|l}
Effective Redshift $z$&Measured $H(z)r_s/{c}$\\
\hline\hline
$z=0.32$&$0.0388\pm 0.0021$\\
\hline
$z=0.57$&$0.0485\pm 0.0013$\\
\hline
$z=2.34$&$0.109\pm 0.002$\\
\hline
\end{tabular}
\caption{Measurements of $H(z)r_s/c$}
\label{htab}
\end{table}

Tension between the $z=2.34$ BAO peak location and the standard cosmological model, at the 2-3$\sigma$ level, was noticed immediately \cite{1404.1801} and has been the subject of numerous investigations.  While there seems to be no standard variation of $\Lambda$CDM that removes this tension \cite{1411.1074}, by combining it with various cosmological datasets it has been noted by several authors that it suggests that the dark energy density becomes negative at high redshift \cite{1404.1801,1405.5116}.  In general the space of parameters is large enough that authors find that this measurement supports models that had previously been focuses of their research, such as modified gravity \cite{1406.2209} or a zero active mass model \cite{1503.05052}.  Needless to say, {\it any} measurement of the dark energy equation of state as a function of redshift $w(z)$ is consistent with an infinite number of dark energy models, such as generalized galileons \cite{gg} and braiding models \cite{bm}.  However, robust evidence that dark energy once contributed negative energy to the universe would imply a conceptual restriction on dark matter models, not just a fitting of parameters.  Therefore it is important to determine just how robust the evidence for a negative energy density really is.

\section{Calculation}

In our analysis we will assume that the universe at large scales is homogeneous and isotropic and is described by Einstein's equations coupled to a perfect fluid with density $\rho$ and pressure $p$, which allow us to define an equation of state $w=p/\rho$.  Note that even many modified gravity models, such as $f(R)$ gravity, can be re-expressed as Einstein gravity coupled to matter \cite{fr,fr2} and so this is quite general.  For simplicity we will now also set the spatial curvature to zero, although the generalization to nonzero curvature is straightforward and later we will argue that to be relevant here the spatial curvature would need to be much larger than current bounds allow.  

In this setting Einstein's equations reduce to Friedmann's equations.  In particular the constraint equation leads to an equation for the Hubble parameter at redshift $z$
\beq
H(z)=\sqrt{\frac{8\pi G\rho(z)}{3}}=\sqrt{\frac{8\pi G}{3}}\sqrt{\rho_M(z)+\rd(z)}
\eeq
where $\rho(z)$, $\rd(z)$ and $\rho_M(z)$ are the total, dark energy and other contributions to the density at redshift $z$ and $G$ is Newton's gravitational constant.  Here we have made the approximation that the spatial curvature is equal to zero.  In the standard cosmological model, $\rho_M(z)$ consists essentially entirely of nonrelativistic matter
\beq
\rho_M(z)=\rho_0(0)(1+z)^3 \label{nonreleq}
\eeq
even if one neutrino flavor is still massless.  The evolution of the dark energy density $\rd$ follows from the Friedmann equations
\beq
\rd(z)=\rd(0)e^{\left(3\int_0^z\frac{1+w(z^\prime)}{1+z^\prime}dz^\prime\right)}.
\eeq
In particular, between the redshifts $z_1$ and $z_2$ the dark energy density evolves as
\beq
\rd(z_2)=\rd(z_1)e^{\left(3\int_{z_1}^{z_2}\frac{1+w(z^\prime)}{1+z^\prime}dz^\prime\right)}.
\eeq

The ratio of two measurements of the Hubble parameter at different redshifts is then
\beq
\left(\frac{H(z_1)}{H(z_2)}\right)^2=\frac{\rho_M(z_1) + \rd(z_1)}{\rho_M(z_2) +\rd(z_1)e^{3\int_{z_1}^{z_2}\frac{1+w(z^\prime)}{1+z^\prime}dz^\prime}} . 
\eeq
If we further make the approximation (\ref{nonreleq}), justified in $\Lambda$CDM, that between redshifts $z_1$ and $z_2$, $\rho_M$ consists entirely of a stable, nonrelativistic perfect fluid we obtain
\beq
\left(\frac{H(z_1)}{H(z_2)}\right)^2=\frac{1 + \frac{\rd(z_1)}{\rho_M(z_1)}}{\left(\frac{1+z_2}{1+z_1}\right)^3 +\frac{\rd(z_1)}{\rho_M(z_1)}e^{3\int_{z_1}^{z_2}\frac{1+w(z^\prime)}{1+z^\prime}dz^\prime}} .  \label{rapeq}
\eeq
The left side is measured using line of sight BAO.  The redshifts $z_1$ and $z_2$ of course are also measured.  Therefore this equation gives a relationship between two unknowns
\beq
\frac{\rd(z_1)}{\rho_M(z_1)}{\rm\ \ and\ \ }e^{3\int_{z_1}^{z_2}\frac{1+w(z^\prime)}{1+z^\prime}dz^\prime}.
\eeq
The first quantity is information about redshift $z_1$ and the second about the evolution between $z_1$ and $z_2$.

Therefore given information about redshift $z_1$, one can learn about the evolution of dark energy between $z_1$ and $z_2$.  This will be the goal of the present note.  As the quantity of interest for dark energy is usually the equation of state $w(z)$, first we will recast this second quantity in a form whose relation to $w(z)$ is more intuitive.



\subsection{Average equation of state}

Although $w(z)$ is an arbitrary function and so can never be fully determined observationally, $H(z)$ only depends on a particular weighted average  (see for example Ref.~\cite{wmap5})
\bea
\wav(z_1,z_2)&=&\frac{\int_{z_1}^{z_2}\frac{1+w(z^\prime)}{1+z^\prime}dz^\prime}{\int_{z_1}^{z_2}\frac{1}{1+z^\prime}dz^\prime}-1\nonumber\\&=&\frac{\int_{z_1}^{z_2}\frac{1+w(z^\prime)}{1+z^\prime}dz^\prime}{{\rm ln}\left(\frac{1+z_2}{1+z_1}\right)}-1
\eea
which can therefore be constrained.

The weighted average $\wav$ has the nice but potentially misleading property that the dark energy density at redshift $z$ is
\beq
\rho(z)=(1+z)^{3(1+\wav(0,z))}\rho(0) \label{waveq}
\eeq
or more generally
\beq
\rho(z_2)=\left(\frac{1+z_2}{1+z_1}\right)^{3(1+\wav(z_1,z_2))}\rho(z_1). 
\eeq
Clearly this is the same formula as for a constant equation of state $w=\wav$.  However we have {\it{not}} assumed that $w$ is constant, even during the redshift interval from $0$ to $z$.  This also does not correspond to a binning approximation in which the redshift is taken to be constant in intervals corresponding to bins.  On the contrary, Eq.~(\ref{waveq}) is exact for any homogeneous evolution of the equation of state $w(z)$, even if there are large fluctuations between redshifts $0$ and $z$.

Substituting Eq.~(\ref{waveq}) into Eq.~(\ref{rapeq}) one finds
\beq
\left(\frac{H(z_1)}{H(z_2)}\right)^2=\frac{\left(\frac{1+z_1}{1+z_2}\right)^3\left(1 + \frac{\rd(z_1)}{\rho_M(z_1)}\right)}{1 +\frac{\rd(z_1)}{\rho_M(z_1)}\left(\frac{1+z_2}{1+z_1}\right)^{3\wav(z_1,z_2)}} .  \label{rapbeq}
\eeq

Now the integrals are gone.  One sees that a measurement of $H(z_1)/H(z_2)$ from a ratio of BAO observations yields a constraint on the two parameters
\beq
\frac{\rd(z_1)}{\rho_M(z_1)}{\rm\ \ and\ \ }\wav(z_1,z_2).
\eeq
Therefore one parameter may be determined in terms of the other.  In particular one can test the $\Lambda$CDM predictions
\beq
\wav=-1,\ \
\frac{\rd(z_1)}{\rho_M(z_1)}=\frac{\left(\frac{1}{\Omega_M}-1\right)}{(1+z_1)^3}
\eeq
where $\Omega_M$ is defined to be
\beq
\Omega_M=\frac{8\pi G\rho_M(0)}{3H_0^2},\ H_0=H(0).
\eeq

In the next subsection we will use this constraint to determine the best fit values $\wav(z_1,z_2)$ given a value of the low-redshift parameter $\rd(z_1)/\rho_M(z_1)$ and no assumptions about redshifts greater than $z_2$.

\subsection{Dark Energy Evolution}


\begin{figure} 
\begin{center}
\includegraphics[width=2.81in,height=1.8in]{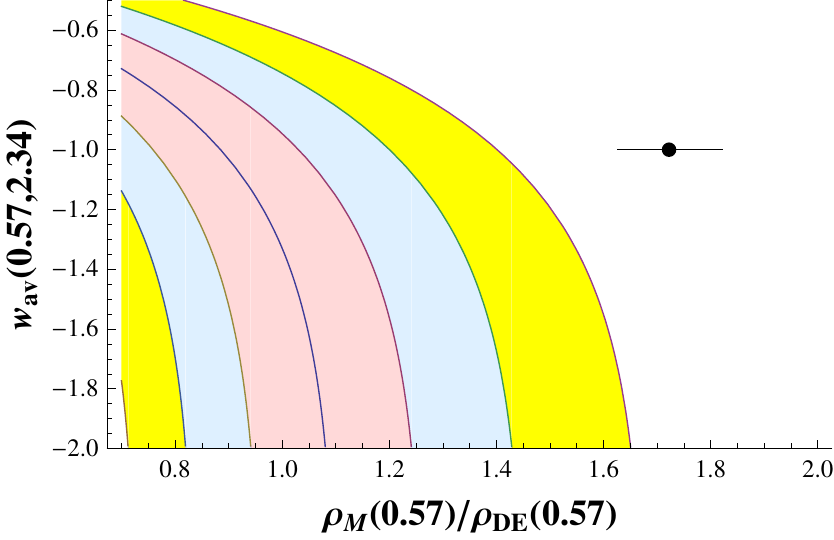}
\caption{The $1\sigma$ (pink), $2\sigma$ (blue) and $3\sigma$ (yellow) allowed intervals for the average dark energy equation of state between redshifts $0.57$ and $2.34$ as a function of the matter to dark energy density ratio at redshift $0.57$.  The circle represents the best fit Planck values \cite{planck}, which are excluded at just over $3\sigma$.  To the right of about 1.1 the preferred $\wav(0.57,2.34)$ goes to negative infinity as $\rd(2.34)$ changes sign, reproducing the negative dark energy solution noted in Refs.~\cite{1404.1801,1405.5116}.}
\label{cfig}
\end{center}
\end{figure}

We will see that the strongest constraint comes from the evolution between $z_1=0.57$ and $z_2=2.34$.  In this case, using the observations from Table~\ref{htab}, Eq.~(\ref{rapbeq}) becomes
\bea
0.198\pm 0.013&=&\left(\frac{H(0.57)}{H(2.34)}\right)^2\label{hvaleq}\\
&=&
\frac{0.104\left(1+\frac{\rd(z_1)}{\rho_M(z_1)}\right)}{1+\frac{\rd(z_1)}{\rho_M(z_1)}9.63^{\wav(z_1,z_2)}}.\nonumber
\eea
Using the Planck best fit value $\Omega_m=0.308\pm 0.012$ from Ref.~\cite{planck}, the $\Lambda$CDM value of the right hand side is $0.155\pm 0.003$.  This indicates about $3\sigma$ of tension with the best fit $\Lambda$CDM model.  In previous analyses \cite{1404.1801} the tension between large $z$ BAO and CMB data is evenly divided by the line of sight BAO and the angular BAO, but here one sees $3\sigma$ arising just from the line of sight BAO even with no prior on the acoustic scale.  This is one of our main results.

In particular, note that the uncertainty is strongly dominated by the BAO measurement.  There is little flexibility in the CMB measurements given $\Lambda$CDM.  Therefore, while it may be tempting to state that this tension may be reduced to a statistically insignificant level by, for example, simply changing $\Omega_m$ to $0.25$, we note that such a small shift is excluded by Planck at nearly the 5$\sigma$ level given $\Lambda$CDM.  {\it The tension can only be resolved within the $\Lambda$CDM framework is if it is caused by a statistical fluctuation or systematic error in the BAO measurements.}

Eq.~(\ref{hvaleq}) is easily solved for $\wav$
\bea
&\wav(0.57,2.34)&\label{princeq}\\&\hspace{-2.1cm}=&\hspace{-2.1cm}\frac{{\rm ln}\left((0.525\pm 0.034)\left(1+\frac{\rho_M(0.57)}{\rd(0.57)}\right)-\frac{\rho_M(0.57)}{\rd(0.57)}\right)}{2.26}.\nonumber
\eea
Note that the $\Lambda$CDM together with Planck's estimate of $\Omega_M$ yield
\beq
\frac{\rho_M(0.57)}{\rd(0.57)}=1.72\pm 0.10
\eeq
and so a complex value of the right hand side, suggesting a complex $\wav(0.57,3.34)$. This reflects the fact that the corresponding best fit $\rd(2.34)$ is negative, as has often been noted in the literature.


\section{Results}
Eq.~(\ref{princeq}) determines the average dark energy equation of state between redshifts $0.57$ and $2.34$ given the ratio of the dark to nonrelativistic energies at $z=0.57$.  No assumptions are made about the universe at $z>2.34$ nor even at $z<0.57$ besides isotropy, homogeneity and zero spatial curvature.  At $0.57<z<2.34$ it is further assumed that all energy except for dark energy has an equation of state $w=0$ and a separately conserved stress energy.  Thus some models in which dark energy and dark matter interact or in which dark matter decays into dark radiation may violate these assumptions if these processes are considerable at $0.57<z<2.34$.  However standard modifications of $\Lambda$CDM such as sterile neutrinos or a running spectral parameter for primordial fluctuations, or even nonstandard modifications such as additional inflationary periods or primordial gravity waves before $z=2.34$, do not affect our bounds.  Furthermore, our bounds are entirely independent of $H_0=H(0)$, and so of the $3\sigma$ tension between supernova and BAO/CMB best fit results.

In Fig.~\ref{cfig} we plot Eq.~(\ref{princeq}) using the crude approximation that the $k\sigma$ uncertainty in the BAO measurement of $H(z)$ is simply $k$ times the $1\sigma$ uncertainty.  The result is compared with the best fit $\Lambda$CDM result from Planck \cite{planck}, represented by a black dot with its error bars.  We note that Fig.~\ref{cfig} looks similar to many plots in the literature in which the allowed range of $w$ is determined in terms of various cosmological parameters within or outside of $\Lambda$CDM.  In those plots generally other parameters, like the number of neutrino flavors and their masses or running spectral index parameters, are held fixed or one marginalizes over them.  We stress that Fig.~\ref{cfig} in this sense is different.  Only the spatial curvature has been arbitrarily fixed.  In that sense we feel that it is unusually robust and model-independent.

One can see that that the Planck value differs from the BAO result by more than 3$\sigma$ and that the Planck/$\Lambda$CDM error bars are smaller.  Therefore while this tension could easily be caused by a statistical fluctuation in BAO, such a possibility is essentially excluded for Planck.  The eBOSS \cite{eboss} and then the DESI experiment will greatly increase the high $z$ BAO sample size \cite{desi} and so can definitively determine whether this tension results from a statistical fluctuation in BAO.  It could in principle also be a systematic error in BAO, however the known systematic effects are very small \cite{review}.

If on the other hand the best fit BAO value is confirmed by future experiments,
what would this imply?  The matter density $\rho_M(0.57)$ is known quite precisely from a variety of distinct measurements, and so it would be difficult to shift sufficiently to reduce this tension.  Thus one may either increase $\rd(0.57)$ or reduce $\wav(0.57,2.34)$.  Without a dramatic shift in $\Omega_M$ and without spatial curvature, an increase in $\rd(0.57)$ requires an increase in $\wav(0,0.57)$.

\begin{figure} 
\begin{center}
\includegraphics[width=2.81in,height=1.8in]{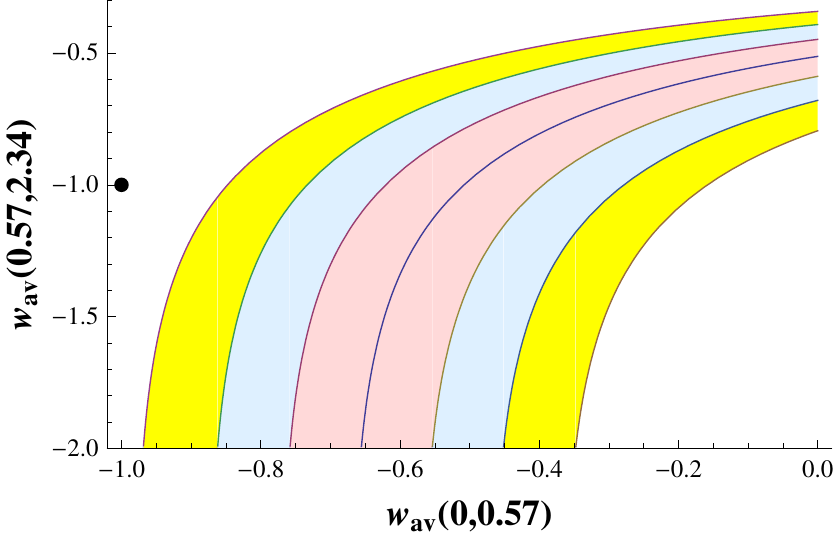}
\caption{The $1\sigma$ (pink), $2\sigma$ (blue) and $3\sigma$ (yellow) allowed intervals for the average dark energy equation of state between redshifts $0.57$ and $2.34$ as a function of the average equation of state between redshifts $0$ and $0.57$, assuming $\Omega_M=0.308$.  The circle represents $\Lambda$CDM.  Note that in the preferred region the average dark energy equation of state decreases with redshift.}
\label{wevfig}
\end{center}
\end{figure}

To be more quantitative, we momentarily assume that that all of the energy density except for dark energy is a nonrelativistic fluid for all redshifts less than $z=2.34$.  In this case
\beq
\frac{\rho_M(0.57)}{\rd(0.57)}=\frac{1.57^{-3\wav(0,0.57)}}{\frac{1}{\Omega_M}-1}.
\eeq
If furthermore we fix $\Omega_M=0.308$ then Eq.~(\ref{princeq}) yields $\wav(0.57,2.34)$ as a function of $\wav(0,0.57)$.  This function is plotted in Fig.~\ref{wevfig}.  From this plot one observes that the dark energy equation of state decreases with increasing redshift.



We note that if one further demands that $\wav(0,0.57)$ is less than $-0.5$, as is indicated for example by type 1a supernova data, then the value of $w(z)$ is greater than -1 at low redshifts and less at higher redshifts, a situation known in the literature as a quintom cosmology \cite{quintom}.  While field theories with standard kinetic terms cannot manifest such behavior \cite{nogo,nogo2}, many newer dark energy models can \cite{sergei,galileon, bm}.

The generalization of this calculation to any two line-of-sight BAO measurements is straightforward.  The other ratios of Hubble scales in Table~\ref{htab} are
\bea
\left(\frac{H(0.32)}{H(0.57)}\right)^2&=&0.64\pm 0.08\\
\left(\frac{H(0.32)}{H(2.34)}\right)^2&=&0.127\pm 0.014.
\eea
One then finds
\bea
&\wav(0.32,0.57)&\\&\hspace{-2.1cm}=&\hspace{-2.1cm}\frac{{\rm ln}\left((0.929\pm 0.112)\left(1+\frac{\rho_M(0.32)}{\rd(0.32)}\right)-\frac{\rho_M(0.32)}{\rd(0.32)}\right)}{0.52}.\nonumber\\
&\wav(0.32,2.34)&\nonumber\\&\hspace{-2.1cm}=&\hspace{-2.1cm}\frac{{\rm ln}\left((0.487\pm 0.056)\left(1+\frac{\rho_M(0.32)}{\rd(0.32)}\right)-\frac{\rho_M(0.32)}{\rd(0.32)}\right)}{2.79}.\nonumber
\eea

\begin{figure} 
\begin{center}
\includegraphics[width=2.81in,height=1.8in]{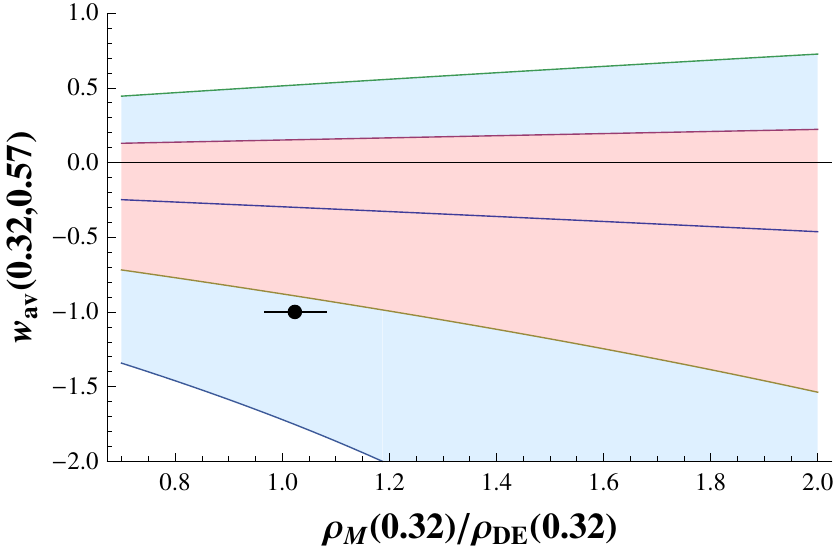}
\includegraphics[width=2.81in,height=1.8in]{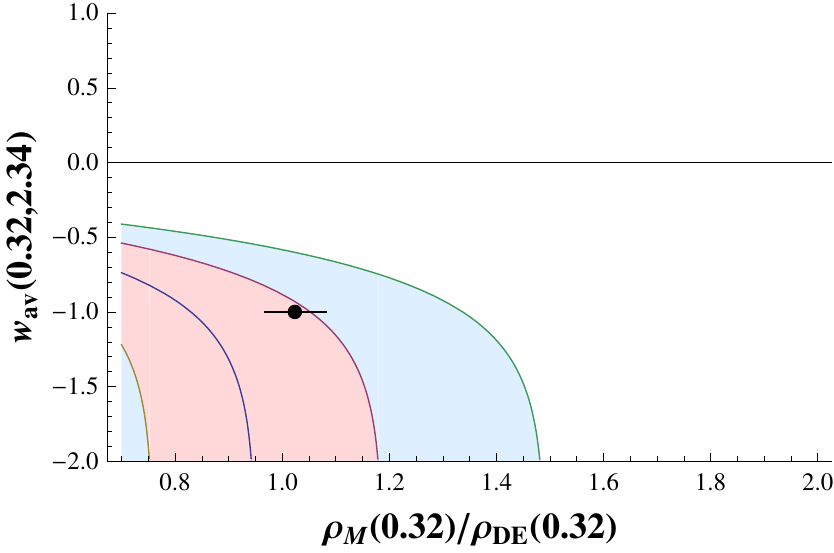}
\caption{The $1\sigma$ (pink) and $2\sigma$ (blue) allowed intervals for the average dark energy equation of state between redshifts $0.32$ and $0.57$ (top) or $2.34$ (bottom) as a function of the matter to dark energy density ratio at redshift $0.32$.  The circle represents the best fit Planck values, which are consistent with the BAO bounds at about the 1$\sigma$ level.}
\label{bfig}
\end{center}
\end{figure}

These two functions are plotted Fig.~\ref{bfig}.  Both are consistent with $\Lambda$CDM at about the $1\sigma$ level.   Apparently the tension arises from higher redshifts.  In the case of the upper panel, which compares $z=0.32$ with $z=0.57$, the constraint is very weak.  On the other hand the lower panel, comparing $z=0.32$ with $z=2.34$ is consistent with $\Lambda$CDM but is quite inconsistent with a model with no dark energy.

\section{Curvature}

Could this tension be resolved by simply adding spatial curvature to the cosmological model?  After all, it has long been appreciated that BAO yields model-independent constraints on curvature \cite{1404.0773}.

Curvature is easily incorporated into Eq.~(\ref{rapbeq})
\bea
&\left(\frac{H(z_1)}{H(z_2)}\right)^2&\\&\hspace{-.5cm}=&\hspace{-1cm}\frac{\left(\frac{1+z_1}{1+z_2}\right)^3\left(1 + \frac{\rd(z_1)}{\rho_M(z_1)}+ \frac{\Omega_K}{\Omega_M (1+z_1)}\right)}{1 +\frac{\rd(z_1)}{\rho_M(z_1)}\left(\frac{1+z_2}{1+z_1}\right)^{3\wav(z_1,z_2)}+\frac{\Omega_K}{\Omega_M (1+z_2)}} \nonumber 
\eea
where
\beq
\Omega_k=\frac{kc^2}{a^2H_0^2}
\eeq
and again we have assumed that the only contributions to the energy density are dark energy and a nonrelativistic fluid for all redshifts less than $z_2$.
If one inserts the measurements of $H(0.57)$ and $H(2.34)$ into this formula and assumes $\Lambda$CDM, together with the Planck value of $\Omega_M$, then one finds $\Omega_k=-0.96$.  Although clearly smaller values of the curvature would be acceptable, yielding less tension, this number is so far beyond current bounds that curvature cannot play a meaningful role in resolving this puzzle.


\section{Conclusions}

We have shown that the full $3\sigma$ of tension results from BOSS line of sight BAO measurements alone, at redshifts $2.34>z>0.57$.  In particular, the dependence upon other datasets is very weak, as the tension may only be relieved with a modification of the current matter density of the universe which is strongly excluded by both Planck and WMAP.  This implies, for example, that this tension cannot be caused by early universe effects such as sterile or unusually massive neutrinos or a running powerlaw of primordial fluctuations, nor by more recent effects such as a shift in the cosmic distance ladder caused by systematic errors in the use of standard candles.  Two possibilities remain.  Either the effect is real, and dark energy evolves, or else the observed Lyman $\alpha$ forest BAO peak has been shifted by statistical or systematic uncertainties.  This second possibility may be excluded in the near future by the next generation of surveys.

\section* {Acknowledgement}

\noindent
JE is supported by NSFC grant 11375201.


\end{document}